\documentclass[preprint,pra,aps,amsmath,amssymb,eqsecnum]{revtex4} %Use this one for preprint format.

\usepackage{color}
\usepackage{graphicx}% Include figure files
\usepackage{dcolumn}% Align table columns on decimal point
\usepackage{bm}% bold math
\usepackage{verbatim}
\usepackage{epsfig}
\usepackage{epstopdf}

\def\id{\mathds{1}}

\def\Re{{\rm Re}}

\begin{document}

%\preprint{Imperial/TP/***}

%\title{Assigning probabilities to particle paths in a double slit experiment using Leggett-Garg inequalities }

%\title{The Double-Slit Experiment as a Leggett-Garg Test}

\title{Leggett-Garg tests for macrorealism: 
interference experiments and the simple harmonic oscillator.}

\author{J.J.Halliwell}

\email{j.halliwell@imperial.ac.uk}

\author{A.Bhatnagar}

\author{E.Ireland}

\author{H.Nadeem}

\author{V.Wimalaweera}

%Wimalaweera, Vinul 
%Bhatnagar, Ansh 
%Nadeem, Hunain 
%Ireland, Ellerey 

\affiliation{Blackett Laboratory \\ Imperial College \\ London SW7
2BZ \\ UK }

\begin{abstract}
%[NOT FOR CIRCULATION].
Leggett-Garg (LG) tests for macrorealism were originally designed to explore quantum coherence on the macroscopic scale. Interference experiments and systems modelled by harmonic oscillators provide useful examples  of situations in which macroscopicity has been approached experimentally and are readily turned into LG tests for a single dichotomic variable $Q$.
Applying this approach to the double-slit experiment in which a non-invasive measurement at the slits is included, we exhibit LG violations. We find that these violations are always accompanied by destructive interference. The converse is not true in general and
we find that there are non-trivial regimes in which there is destructive interference but the two-time LG inequalities are satisfied which implies that it is in fact often possible to assign (indirectly determined) probabilities for the interferometer paths. Similar features have been observed in recent work involving a LG analysis of a Mach-Zehnder interferometer and we compare with those results.
We extend the analysis to the triple-slit experiment again finding LG violations,
and we also exhibit examples of some surprising relationships between LG inequalities and NSIT conditions that do not exist for dichotomic variables. For the simple harmonic oscillator, we find an analytically tractable example showing a two-time LG violation with a gaussian initial state, echoing in simpler form recent results of Bose et al (Phys. Rev. Lett. 120, 210402 (2018)).
\end{abstract}

%but to date their application to macroscopic systems remains a distant goal. Interference experiments provide a class of interesting  situations in which macroscopicity is being approached experimentally and may therefore also be fruitful ground for Leggett-Garg tests. Here, we show how the traditional double-slit experiment can be employed as a Leggett-Garg test.  This approach provides an indirect way of assessing whether there exist probabilities for the particle paths which match the marginal probabilities at both the slits and at the screen. These probabilities exist when a set of LG inequalities are satisfied, which is readily achievable for non-trivial parameter ranges, perhaps contrary to traditional lore about the double-slit experiment. LG violations indicate the absence of probabilities for the paths and these violations are shown to occur precisely when the interference pattern shows destructive interference, thus showing a coincidence of two very different notions of quantum-mechanical behaviour.
%\end{abstract}

%This probabilities, when they exist, may be determined indirectly using sets of different experiments, or directly, using a pair of sequential measurements in which the first one is weak. We find non-trivial parameter ranges in which the LG inequalities are both satisfied or violated and we also show that these ranges correspond precisely to situations in which the interference pattern on the screen displaces constructive interference or destructive interference.
%\end{abstract}

%\pacs{03.65.Ta, 03.65.Xp}

\maketitle

%\centerline{Imperial/TP/03-4/7}

% Some convenient definitions below

\newcommand\beq{\begin{equation}}
\newcommand\eeq{\end{equation}}
\newcommand\bea{\begin{eqnarray}}
\newcommand\eea{\end{eqnarray}}

\def\A{{\cal A}}
\def\D{\Delta}
\def\H{{\cal H}}
\def\E{{\cal E}}
\def\p{\partial}
\def\la{\langle}
\def\ra{\rangle}
\def\ria{\rightarrow}
\def\x{{\bf x}}
\def\y{{\bf y}}
\def\k{{\bf k}}
\def\q{{\bf q}}
\def\p{{\bf p}}
\def\P{{\bf P}}
\def\r{{\bf r}}
\def\s{{\sigma}}
\def\a{\alpha}
\def\b{\beta}
\def\e{\epsilon}
\def\U{\Upsilon}
\def\G{\Gamma}
\def\om{{\omega}}
\def\Tr{{\rm Tr}}
\def\ih{{ \frac {i} { \hbar} }}
\def\trho{{\rho}}
\def\au{{\underline \alpha}}
\def\bu{{\underline \beta}}
\def\pp{{\prime\prime}}
\def\id{{1 \!\! 1 }}
\def\half{\frac {1} {2}}
\def\jjh{j.halliwell@ic.ac.uk}

\section{Introduction}

The Leggett-Garg (LG) inequalities were proposed in order to test the world view known as macrorealism, the view that a macroscopic object evolving in time may possess definite properties independent of past or future measurements \cite{LeGa,L1}. 
The main motive for developing such tests is that they offer the possibility of assessing whether macrosopic objects may exist in coherent superpositions. An affirmative answer to this question would rule out interesting families of alternatives to quantum theory in which macroscopic superpositions are suppressed \cite{MaTi}.

LG tests typically concern measurements of a single dichotomic variable $Q$ in experiments involving either single times or pairs of times thereby determining the averages $ \langle Q_i \rangle $ and correlators $C_{ij} = \langle Q_i Q_j \rangle$, where $Q_i $ denotes $Q(t_i)$ and $i,j=1,2,3$. Macrorealism is characterized by three requirements: (i) macrorealism per se (the variables take definite values); (ii) non-invasive measurability (the present state cannot be affected by a past measurement), (iii) induction (future measurements cannot affect the present state). These assumptions ensure the existence of a joint probability distribution at three times on the variables $Q_i$ which in turn implies that the averages and correlators obey the following two types of inequality: the three-time LG inequalities,
\beq
1 + C_{12} + C_{23} + C_{13} \ge 0,
\label{LG3s}
\eeq
plus the three more obtained by flipping the sign of each $Q_i$, and the two-time LG inequalities,
\beq 1+ \langle Q_i \rangle + \langle Q_j \rangle + C_{ij} \ge 0,
\label{LG2s}
\eeq
plus three more from the same sign flips, where $ij = 12, 23, 13$.
For a three-time situation there are a total of four three-time LG inequalities and twelve two-time inequalities. This set of sixteen form a set of conditions for macrorealism which are both necessary and sufficient \cite{HalLG4}. However, purely for experimental convenience it is often simpler to work solely with the two-time LG inequalities and indeed many of the experiments testing the LG inequalites do precisely this. Necessary and sufficient conditions of this general form have also been found for multi-time measurements \cite{HaMa} and many-valued variables \cite{HaMa2}.

There have been considerable theoretical developments of the LG framework over the years and numerous experimental tests (see Ref.\cite{ELN} for a useful review), and they typically refute a macrorealistic description of nature. Some such tests have come close to genuine 
macroscopicity \cite{Knee0} but it would clearly be of interest to find more experimental situations in which this can be achieved.
%and few approach macroscopicity
%truly macroscopic experimental test has yet to be performed. 
One way to proceed would be to examine different types of experiments which were not originally designed to be LG tests but were designed to explore macroscopic quantum effects. 
Two classes of systems naturally fit the bill. First, as noted by Pan \cite{Pan}, interference experiments of the double-slit variety have now been developed to the point that interference effects can be detected for sizeable objects, with reasonable claim of macroscopicity \cite{Arndt}. Second, a number of recent experiments have shown it is possible to trap reasonably large masses in harmonic wells \cite{Bose}.
Hence it is clearly of considerable interest to develop tests of macrorealism for both interference experiments of the double-slit type and simple harmonic oscillators. 

The primary aim of this paper is to explore LG and related tests for macrorealism for the double-slit experiment, the triple-slit experiment and the simple harmonic oscillator. In each case we set up the requisite formalism and exhibit parameter ranges in which various macrorealistic conditions are violated or satisfied. A secondary aim, 
for the double-slit and triple-slit experiments, is to explore the relationship between the LG inequalities and destructive interference.
%When the LG inequalities are satisfied one can in effect assig
%We also explore the relationship between LG violation and destructive interference in the double slit experiment and shed some light on the old question as to the degree to which one can assign probabilities to particle paths.

We begin in Section 2 with a presentation of the general LG framework for measurements at two times, modestly adapted (along the lines discussed in Ref.\cite{HaMa2}) for the traditional 
double-slit experiment from the perspective of the LG framework. 
The detailed application to the double-slit experiment is given in Section 3. We readily find parameter ranges for which the LG inequalites are violated or satisfied.
We then examine the relationship between the Leggett-Garg inequalities for the particle paths and the interference pattern.
We find that LG violations are always accompanied by destructive interference, hence these two quite different notions of quantumness are related. But the converse is not true and we find significant regimes in which there is destructive interference but the LG inequalities are satistied. This sheds some light on the old question as to the degree to which one can assign probabilities to the particle paths in an interference experiment when destructive interference is present -- it is in fact possible for a set of indirectly constructed probabilities.
Our results have some overlap with recent work on LG tests using a Mach-Zehnder interferometer 
discussed in Refs. \cite{Pan,KoBr}. (A number of possible or actual experiments in this area have also been discussed \cite{Robens,Asa,Myung}.)
%\tb{TESTS?}

In Section 4 we extend our considerations to the triple-slit experiment which has some interesting new features. 
We again readily find regimes in which the LG inequalities are satisfied or violated.
However, we find the relationship between destructive interference and LG violation is more fluid, with clear logical relations between them only existing for specific parameter values.
Of perhaps greater interest is that the triple-slit experiment provides an arena in which some of the macrorealism conditions for many-valued variables proposed in Ref.\cite{HaMa2}
can be put to experimental test. In particular, this system shows a non-trivial relationship between LG violations and no-signaling in time (NSIT) conditions (unlike systems with dichotomic variables for which the latter simply implies the former) and some of these relationships are determined.

%\tb{We briefly discuss possible violations of the L\"uders bound (the maximal LG violation allowable by quantum mechanics for measurements of dichotomic variables, numerically the same as the Tsirelson bound \cite{Tsi} in Bell experiments).}

In Section 5, turning to the different territory of a simple harmonic oscillator,
we investigate a set of two-time LG inequalities with the initial state a coherent state, a simpler version of the general programme initiated by Bose et al
\cite{Bose}. We find exact analytic expressions for the averages and single correlator 
(in contrast to the numerical results of Bose et al)
and exhibit a regime in which LG violation is possible. We identify the origin of the LG violation in this case, 
using Wigner function language, 
and contrast with that arising in interference experiments.
This part of our work is also a natural progression from a recent work in which a LG analysis of the free particle and the arrival time problem was carried out \cite{HalTOA}.
We summarize and conclude in Section 6. 
%We also highlight specific points that may be of interest for experiment test.
% SHO example

\section{General Analysis of Measurements at Two Times for the Double Slit Experiment}

\begin{figure}
  \centering
  \includegraphics[width=.6\linewidth]{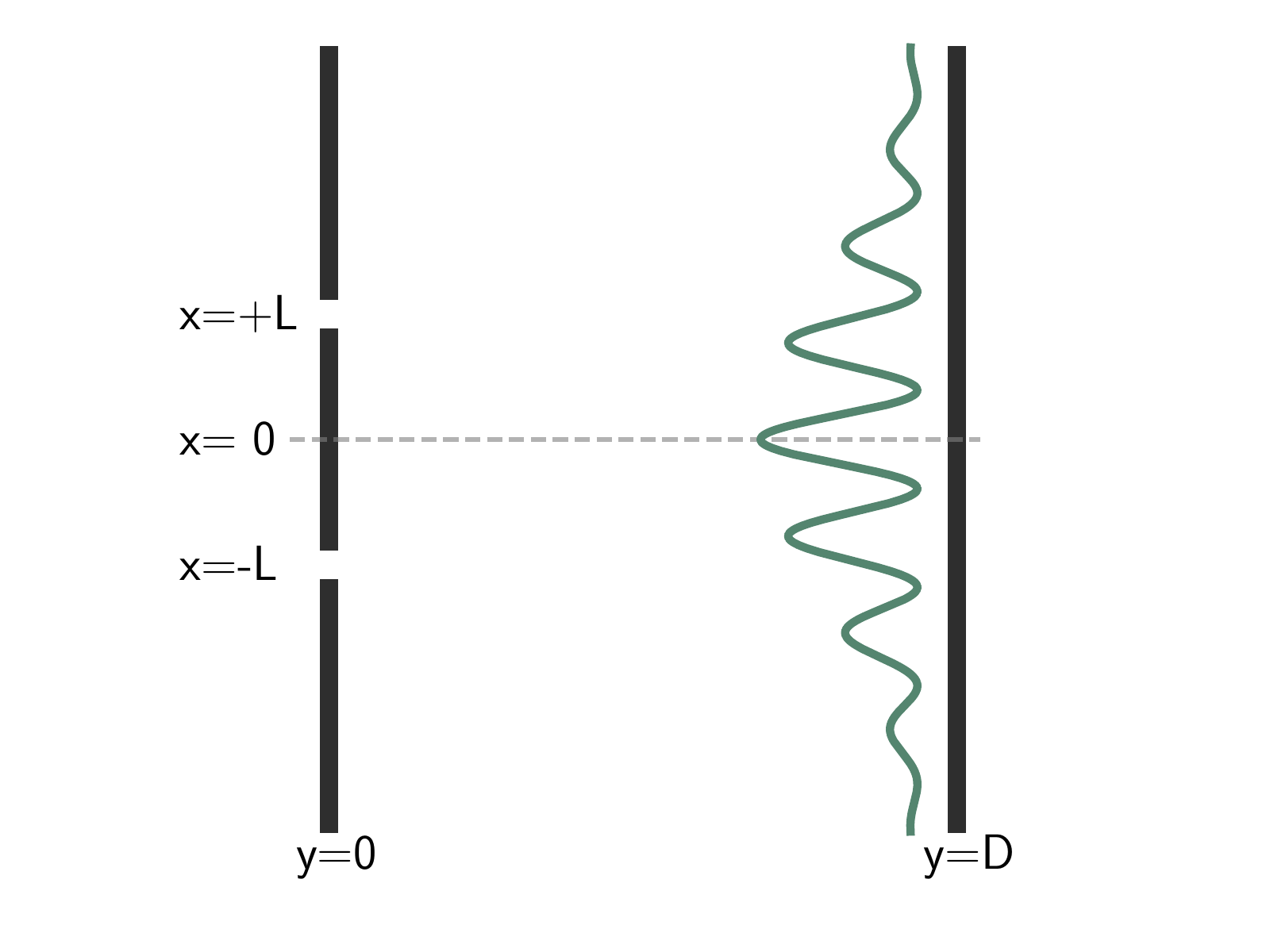}
  \caption{The experimental setup for the double-slit. The slits are centered at $x=\pm L$ and
  $y=0$.  The particle propagates from the slit to the screen at $y=D$ where the interference
  pattern is measured.}
  \label{fig:double_slit_figure}
\end{figure}

We begin with a quantum-mechanical description of the double slit experiment. We consider motion in the $xy$ plane with the slits and screen taken to be lines of constant $y$ and consider an incoming state $\rho$  approaching the slits in the $y$ direction. See Figure \ref{fig:double_slit_figure}.

At $t_1$ there is a projective measurement described by projection operator $P_s$ onto values $s=\pm 1$ denoting which slit the particle went through (which could simply be projections on to the positive or negative $x$-axis)
and followed by time evolution to time $t_2$ and second projective measurement $E_n$ onto one of many values $n$, denoting a coarse-grained measurement of the screen position. We write this projector
\beq
E_n = \int_{\Delta_n} dx \ |x \rangle \langle x |.
\eeq
Here, $\Delta_n$ denotes one of many small intervals which divide up the real line, each of size $\Delta$. Generally each $\Delta_n$ is taken to be arbitarily small but may be kept for normalization purposes. (Following Ref.\cite{HaMa2}, we use $P_s$ for projections onto dichotomic variables and $E_n$ for projections onto variables with three or more values).

The two-time probability for these sequential measurements is,
\beq
p_{12}(s_1, n_2) = {\rm Tr} \left( E_{n_2} (t_2)  P_{s_1}  (t_1) \rho P_{s_1} (t_1)  \right).
\label{p12}
\eeq
(The formalism given here follows Ref.\cite{HalLG1}).
It matches the single time probability $p_1 (s_1) = {\rm Tr} ( P_{s_1} (t_1)  \rho )$ when summed over $n_2$. The interference pattern at the screen is given by the probability
\beq
p_2 (n_2) = {\rm Tr} ( E_{n_2} (t_2) \rho),
\label{p2}
\eeq
in which there is no earlier measurement at $t_1$. (Here the subscripts on the probabilities denote the quantities measured). It does not coincide with Eq.(\ref{p12}) summed over $s_1$, i.e. the so-called no-signaling in time (NSIT) condition \cite{KoBr,Cle}
\beq
p_2 (n_2) = \sum_{s_1} p_{12}(s_1, n_2),
\label{NSIT}
\eeq
is not satisfied in general.
This is the familiar fact that sequentially measured probabilities for pairs of paths in the double slit experiment suffer from interference and fail to satisfy the probability sum rules.

Here, however, we are concerned with looking for other, indirect procedures for determining possible probabilities for the paths in a double slit experiment. In particular, we consider the quasi-probability,
\beq
q(s_1, n_2) = {\rm Re} {\rm Tr} \left(  E_{n_2} (t_2) P_{s_1} (t_1)  \rho  \right).
\label{quasi}
\eeq
(See Refs.\cite{HalLG1,GoPa,HaYe2}).
This quantity does match the two marginals, $p_1 (s_1)$ and $p_2 (n_2)$, and in particular, we have that
\beq
p_2 (n_2) = \sum_{s_1} q(s_1, n_2),
\label{sumq1}
\eeq
in contrast to Eq.(\ref{p12}). However, the quasi-probability can be negative, which we regard as an indication of quantum-mechanical behaviour.
When non-negative, it is our candidate expression for the path probabilities in the double-slit experiment and Eq.(\ref{sumq1}) is then the statement that the interference pattern probability {\it can} be regarded as the sum of two path probabilities.

The quasi-probability Eq.(\ref{quasi})  has a simple relation to the standard quantum-mechanical two-time probability Eq.(\ref{p12}), namely,
\beq
q(s_1, n_2) = p_{12} (s_1, n_2)  + {\rm  Re} D (s_1,n_2 |-s_1,n_2),
\label{qp}
\eeq
where the quantity
\beq
D (s_1,n_2 |s_1',n_2) = {\rm Tr} \left(  E_{n_2} (t_2) P_{s_1} (t_1) \rho P_{s_1'} (t_1)  \right),
\label{DF}
\eeq
is the decoherence functional, and is a measure of the interference between different paths in the interferometer. (Here we employ the mathematical machinery of the decoherent histories approach to quantum theory \cite{HalLG1,deco}).
Note that this interference term vanishes when the NSIT condtion Eq.(\ref{NSIT}) holds.
By summing Eq.(\ref{qp}) over $s_1$, we obtain
\beq
p_2 (n_2) = \sum_{s_1} p_{12} (s_1,n_2) +  2 \ {\rm  Re} D (s_1,n_2 |-s_1,n_2).
\label{sumq}
\eeq
(Noting that ${\rm  Re} D (s_1,n_2 |-s_1,n_2)$ is in fact independent of $s_1$, since $s_1$ takes only values $\pm 1$).
This relation shows precisely how non-zero interference prevents the sum rules from being satisfied.  

Furthermore, one can see immediately how the negativity of the quasi-probability may be related to the interference pattern. The right-hand side of Eq.(\ref{sumq}) consists of a positive term (essentially the mean size of the interference pattern) plus an interference term which can be positive (constructive interference) or negative (destructive interference). From Eq.(\ref{qp}), one can see that if the quasi-probability is negative then $ {\rm  Re} D (s_1,n_2 |-s_1,n_2) $ must be negative and sufficiently large. Hence we see in quite a general way, beyond the specifics of particular interference experiments, that {\it negative quasi-probability implies destructive interference}.

We now sketch the relationship between the quasi-probability and the LG inequalities 
(given earlier, for example, in Ref.\cite{HalLG1}).
For simplicity, we first consider the case in which the projections at both times are two-valued and take the form $P_s = (1 + s \hat Q )/2$, for a dichotomic variable $\hat Q$. (We use hats to denote operators only when the difference between classical and quantum versions is not obvious, which in practice affects only the variable $Q$).
The quasi-probability is then conveniently expanded in terms of its moments,
\beq
q(s_1,s_2) = \frac {1}{4} \left(1 + \langle \hat Q_1 \rangle  s_1 +  \langle \hat Q_2 \rangle s_2  + C_{12} s_1 s_2 \right),
\label{mom}
\eeq
where the correlation function $C_{12}$ is given by,
\beq
C_{12} = \half \langle \hat Q_1 \hat Q_2 + \hat Q_2 \hat Q_1 \rangle,
\label{corr}
\eeq
and we use the notation $\hat Q_1 = \hat Q(t_1), \hat Q_2  = \hat Q (t_2) $. The conditions
\beq
q(s_1, s_2) \ge 0,
\label{qss}
\eeq
are then precisely a set of four two-time LG inequalities, Eq.(\ref{LG2s}),  written in quantum form. Macrorealistically, they correspond to the situation in which we make measurements of variables $Q_1$ and $Q_2$ at times $t_1$ and $t_2$ to determine the averages $\langle Q_1 \rangle$, $\langle Q_2 \rangle$ and the correlator $C_{12} = \langle Q_1 Q_2 \rangle$. In quantum theory,
the inequality Eq.(\ref{qss}) can be violated up to a maximum of $ - \frac{1}{8}$, which corresponds to a violation of $ -\frac{1}{2}$ in the LG inequalities, the so-called L\"uders bound (proved for example in Ref.\cite{HalTOA}).

%For a macrorealistic theory, a joint probability for $Q_1$ and $Q_2$ exists which implies that the inequalities
%\beq
%\langle  (1 + s_1 Q_1 ) (1 + s_2 Q_2 ) \rangle \ge 0,
%\eeq
%must hold. Expanded out, this yields the LG inequalities,
%\beq
%1 + s_1 \langle Q_1 \rangle + s_2 \langle Q_2 \rangle + s_1 s_2 C_{12} \ge 0.
%\label{LG2}
%\eeq
%These are clearly necessary conditions for macrorealism. They are also sufficient since the inequalities themselves, when satisfied and multiplied by $\frac{1}{4}$, are the probabilities for the two histories matching the measured data.

The LG inequalities Eq.(\ref{LG2s}) define a version of macrorealism christened weak macrorealism in Ref.\cite{HalLG4}. This is to contrast it with alternative definitions characterized by the NSIT condition Eq.(\ref{NSIT}) being satisfied, which is referred in Ref.\cite{HalLG4} to as strong  macrorealism. These two conditions have a clear logical relationship, namely that the NSIT condition implies the LG inequalties but not conversely.
However, this logical relationship no longer holds when we go beyond dichotomic variables \cite{HaMa2}, as is the case in the triple-slit experiment considered below.

Returning to the case in which measurements are made at the screen using the many-valued projector $E_n$, we can easily relate this to the dichotomic case by picking a fixed value of $n$ and defining a dichotomic variable $Q(n) = 2E_n - \id$, where $\id$ is the identity operator. It is then easily seen that the two-time LG inequalities for each $Q(n)$ are equivalent to the condition
\beq
q(s_1, n_2) \ge 0.
\eeq
on the quasi-probability Eq.(\ref{quasi}). This set of relations is therefore the natural generalization of the standard LG inequalities, generalized to many-valued variables at the second time \cite{HaMa2}. 
%We also see from this the clear relationships between destructive interference and LG inequality violation. 

Turning now to measurement procedures, the standard two-time LG inequalities are tested by measuring 
$\langle Q_1 \rangle$, $\langle Q_2 \rangle$ and the correlator $C_{12} $ in three different experiments, where the experiment measuring the correlator must be done non-invasively \cite{HalLG4,Knee,Rob,Kat,HalNIM}. This is typically achieved using ideal negative measurements but other methods exist (see for example Refs. \cite{HalLG3,Maj,MajThe}).
However, rather than measuring a set of dichotomic variables at the screen, there is a more convenient way.
This is to note from Eqs.(\ref{qp}), (\ref{sumq}), that the interference term may be eliminated and we find the formula,
\beq
q(s_1, n_2) = p_{12} (s_1, n_2) + \frac{1}{2} \left( p_2 (n_2) - \sum_{s_1'} p_{12} (s_1',n_2) \right).
\label{qMR}
\eeq
This quantity can therefore be determined by measuring $p_{12} (s_1, n_2)$ using an ideal negative measurement and measuring $p_{2} (n_2)$ in a separate experiment. 
Furthermore, although this formula is derived using quantum mechanics, it can be postulated as a candidate probability in purely macrorealistic terms. 
%For suppose we first meaure $p_{12} (s_1, n_2)$ and discover it fails to satisfy the sum rules, i.e. to match $p_2 (n_2)$ when summed over $s_1$. One could simply then try to modify the sequentially measured formula by a term proportional to the sum rule violation, in such a way as to ensure that both marginals $p_1 (s_1)$ and $p_2 (n_2)$ can be matched. Eq.(\ref{qMR}) is the obvious guess through which this may be accomplished.
This method of determining $q(s_1, n_2)$ is in practice probably the easiest to implement experimentally, since as we shall see in the next section, it can avoid the issue of overall normalization.

The quasi-probability $q(s_1, n_2)$ also naturally arises in the context of pairs of sequential measurements in which the first one is weak \cite{HalLG1} (although reservations have been expressed as to whether such measurements meet the non-invasiveness requirement \cite{ELN,Wilde}). We note also an interesting connection between the two-time LG inequalities (and hence the above quasi-probability) and anomalous weak values  \cite{Pan,Dressel}.

\section{Explicit Calculation for the Double Slit Experiment}

We now apply the formalism to developed in the previous section for specific initial states for the double-slit experiment.
We choose an initial state
\beq
| \psi \rangle = \sum_s \alpha_s | \psi_s \rangle,
\eeq
which represents the state immediately after the particle has impinged on the slits, where $ \sum_s | \alpha_s |^2= 1$.
The detailed form of the states $| \psi_{\pm} \rangle $ is not required -- we assume only that they strongly concentrated at $x = \pm L$,
%\beq
%\psi_{\pm} (x) = \frac{1} { (2 \pi \sigma^2)^{\frac{1}{4} } } \exp \left( - \frac { (x \mp L)^2 }{ 4 \sigma^2} \right) 
%\label{gauss}
%\eeq
and are approximate eigenstates of $P_{s_1}$.
We readily find
\bea
q(s_1, n_2) &=& {\rm Re} \sum_{s} \alpha_s^* \alpha_{s_1} \langle \psi_{s} (\tau) | E_{n_2} | \psi_{s_1} (\tau) \rangle
\nonumber \\
&=& | \alpha_{s_1} |^2  \langle \psi_{s_1} (\tau) | E_{n_2} | \psi_{s_1} (\tau) \rangle
+ {\rm Re} \sum_{s \ne s_1} \alpha_s^* \alpha_{s_1} \langle \psi_{s} (\tau) | E_{n_2} | \psi_{s_1} (\tau) \rangle,
\eea
where for convenience we take $t_1 = 0$ and $t_2 = \tau$, and note that the time-dependence has been shifted from the projector to the initial state.
(There is only one term in the sum, namely $s = -s_1$, but the form given generalizes to the three-slit case).
We also find,
\bea
p_{12} (s_1, n_2) &=&  | \alpha_{s_1} |^2  \langle \psi_{s_1} (\tau) | E_{n_2} | \psi_{s_1} (\tau) \rangle,
\\
p_2 (n_2) &=& \sum_{s, s_1} \alpha_s^* \alpha_{s_1} \langle \psi_{s} (\tau) | E_{n_2} | \psi_{s_1} (\tau).
\eea
In all these expressions, $|\psi (\tau) \rangle = \exp( - \frac{i}{\hbar} H \tau ) | \psi \rangle$, where $H$ is the free particle Hamiltonian.

We now make the simplification that the projections at the screen are onto extremely narrow ranges of size $\Delta$ so that $E_n$ is approximated  by $ \Delta  |x \rangle \langle x|$ where $ x = n \Delta $.
This means that
\beq
\langle \psi_{s} (\tau) | E_{n_2} | \psi_{s_1} (\tau) \rangle \approx \Delta \ \psi^*_s (x,\tau) \psi_{s_1} (x, \tau).
\eeq
Approximating the initial wave functions $\psi_s (x)$ by $\delta$-functions $ \delta (x - s L )$, up to normalization, and evolving in time, we find
\beq
\psi_s (x, \tau) \approx N_{\tau} \exp \left( i \frac{m ( x-s L)^2 }{ 2 \hbar \tau} \right).
\eeq
The prefactor  factor $N_{\tau} $ will slowly decay to zero for large $|x|$ in a more exact treament but the explicit form is not required in what follows.
%[MORE DETAILED CALCULATION NEEDED HERE FOR GAUSSIANS].
We also take
\beq
\alpha_+ = \cos \phi \ e^{ i \theta_+}, \ \ \ \alpha_- =  \sin \phi \ e^{ i \theta_-},
\eeq
and define $\theta = \theta_+ - \theta_-$. It is convenient to define quasi-probability densities $q(s_1, x)  = q(s_1, n_2) /\Delta $  (estimating the risk of notational confusion to be small), and we get
\bea
q(+,x) &=& \frac{1}{2} |N_t|^2 \left( 1 +  \cos 2 \phi + \sin 2 \phi \cos \left( \frac {2 m x L} { \hbar \tau} - \theta \right) \right),
\\
q(-,x) &=& \frac{1}{2} |N_t|^2 \left( 1 -  \cos 2 \phi + \sin 2 \phi \cos \left( \frac {2 m x L} { \hbar \tau} - \theta \right) \right).
\eea
It is readily seen that there are substantial parameter ranges for which either of these quasi-probabilities are negative. 
For example $ \phi = 5 \pi / 8$ and $x=0$, $\theta = 0$ yields $q(+,x) =  |N_t|^2 (1 - \sqrt{2})/ 2 < 0 $. Hence there is a clear LG violation.

%\tb{DETAILS?}

The screen probability density is
\bea
p_2 (x) &=& q(-,x) + q(+,x) 
\\ 
&=& |N_t|^2 \left( 1  + \sin 2 \phi \cos \left( \frac {2 m x L} { \hbar \tau} - \theta \right) \right).
\label{p2x}
\eea
We also get
\beq
p_{12} (s_1, x ) = \frac{1}{2} | N_t |^2 \left( 1 + s_1 \cos 2 \phi \right),
\eeq
and the mean of the interference pattern is,
\beq
\sum_{s_1} p_{12} (s_1, x) = | N_t|^2.
\eeq
These relations are all consistent with Eq.(\ref{sumq}), as expected.

\begin{figure}
  \centering
  \includegraphics[width=.7\linewidth]{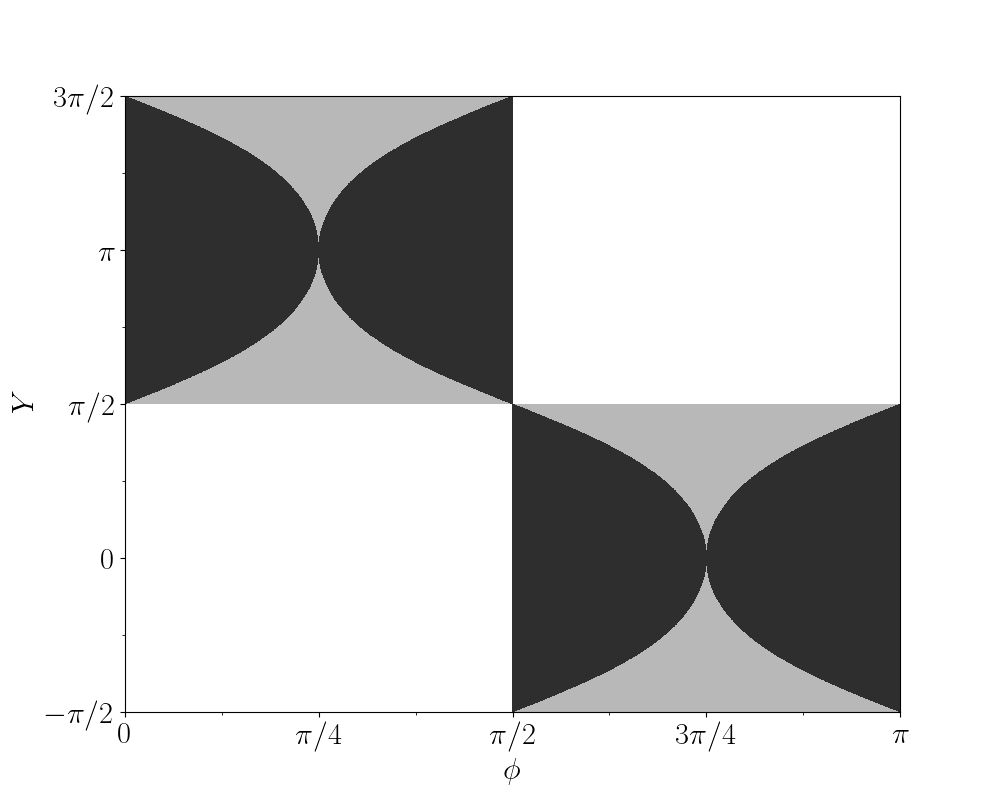}
  \caption{The grey/black rectangles show the regions in the parameter space $(\phi, Y)$ for which there is destructive interference (and no destructive interference in the white rectangles).
In the black regions there is also LG violation, but no violation in the grey regions. The figure clearly shows that regions of LG violation correspond to destructive interference but destructive interference can occur without LG violation for a wide range of parameters. LG violation and destructive interference can however coincide for special parameter choices.}
%One can also see that for values of $\phi$ close to $ n \pi/ 2$, for $n=0, \pm 1, \pm 2 $, there is an LG violation except for a very small range of values of $y$.
  \label{fig:quasi_destructive_double_slit}
\end{figure}

To explore the connection between the sign of the quasi-probabilities and the interference pattern, first note that only one of $q(+,x)$ and $q(-,x)$ can  be negative, since they sum to a positive number. It is then
convenient to examine the quantity,
\beq
q(+,x) q(-,x) = \frac{1}{4} |N_t|^4 \left[ \left( 1  + \sin 2 \phi \cos \left( \frac {2 m x L} { \hbar \tau} - \theta \right) \right)^2 - \cos^2 2 \phi \right],
\label{qq}
\eeq
which is negative if either one is negative. Comparing with Eq.(\ref{p2x}), it is then clear that the only way for either quasi-probability to be negative is for the interference pattern to show destructive interference, i.e. the interference term in Eq.(\ref{p2x}) is negative, that is
\beq
\sin 2 \phi \cos Y  < 0,
\eeq
where
\beq
Y = \frac {2 m x L} { \hbar \tau} - \theta.
\label{ydef}
\eeq
Plots of the parameter ranges for which there is destructive interference and LG violation are shown in  Figure \ref{fig:quasi_destructive_double_slit}.

\begin{figure}
  \centering
  \includegraphics[width=.7\linewidth]{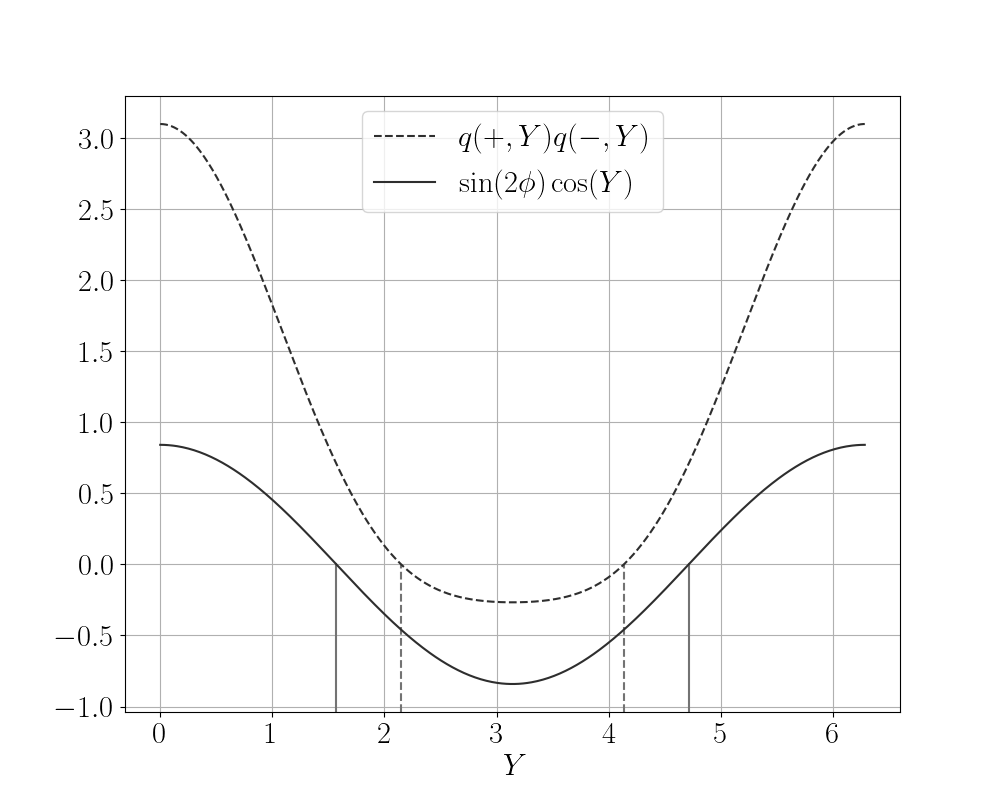}
  \caption{A plot of the LG violation discriminant Eq.(\ref{qq}) as a function of $Y$ (up to an overall factor) and of the size of the interference term $ \sin (2 \phi) \cos (Y) $ in Eq.(\ref{p2x}), both for $ \phi = 0.5$ (a typical value). 
There are clear ranges of $Y$ (between the pairs of vertical lines) for which there is destructive interference but no LG violation.}
\label{fig3}
\end{figure}

%\begin{figure}
 % \centering
 % \includegraphics[width=.7\linewidth]{fig4}
  %\caption{A plot of the LG violation discriminant Eq.(\ref{qq}) and the size of the interference term in Eq.(\ref{p2x}) for the special value $\phi = \pi/4$. The destructive interference reaches its maximum value but the LG inequalities are always satisfied for all $Y$. }
%\label{fig4}
%\end{figure}

The plots show the following features. The most important feature is that LG violation,
negativity of Eq.(\ref{qq}), is always accompanied by destructive interference. But the converse is not true in general -- there are significant regions where there is destructive interference but no LG violation. This means that it is in fact often possible to assign probabilities to the paths in a double-slit experiment even in the face of destructive interference, perhaps contrary to popular lore.
This feature is also shown plotted differently in Figure \ref{fig3}.
Note however, that these probabilities are assigned by an indirect procedure in which they are assembled from a set of different measurements. Note also that this indirect procedure is different from the study of the double-slit experiment, in, for example, Refs.\cite{WoZu,Koc}, in which a variable strength measurement at the slits is employed, and the resulting diminishment of the interference pattern observed. A closely related quasi-probability can be obtained in this way \cite{HalTOA} but this is not a LG test due to the invasiveness of the measurement, even if very weak.

There are special parameter values for which LG violation and destructive interference are more tightly related. First,
for values of $\phi$ in the neighbourhood of $ n \pi / 2 $, for $n=0,\pm 1, \pm 2 $, the regions of destructive interference with no LG violation are very small. Hence there is an approximate coincidence. (Although note that both the quasi-probabilities and interferences are close to zero for these values.)

Second, for $Y=0$ and $Y = \pi$, there are significant ranges of $\phi$ for which destructive interference coincides entirely with LG violation. The analysis here of LG violations and destructive interference is then in fact mathematically the same as the analysis of interferences in the MZ inteferometer considered by Pan \cite{Pan} who found a perfect coincidence between LG violations and destructive interference. 

%This coincidence of course follows from a quantum-mechanical argument but a plausible macrorealistic argument that destructive interference implies LG violation in this case was also given. This was then argued to imply that the non-invasiveness assumption made in LG tests is not in fact required, since the LG violation is deduced indirectly from a single final measurement. This is an appealing conclusion which avoids one of the most fraught issues of LG tests. Here, this argument only applies for very specific parameter values so we will stay with the usual assumption that a non-invasive measurement is performed at the first time in order to check for LG violations.

A similar LG analysis of the MZ interferometer was also undertaken by Kofler and Brukner \cite{KoBr} who came to rather different conclusions to those obtained here and by Pan. They also state that there seemed to be no violable LG inequality for the double-slit experiment. However, their conclusions are based on a three-time LG inequality, not the two-time one used here, so we find no contradiction with their conclusions.

%For all but a discrete set of values of $\phi$ there is essentially always some value of $y$ for which one of the LG inequalities is violated.

%The degree of LG violation and destructive interference are not simply related. The maximum LG violation occurs when the destructive interference is at about half its maximum. On the other hand 
%there are discrete values of $\phi$ for which the LG inequalities are satisfied for all $Y$ but these in fact correspond to
%to maximal destructive interference. For example, this happens at $\phi = \pi/4$ and $Y = \pi$ and this is shown more explicitly in Figure \ref{fig4}. 
%The two quasi-probabilities are actually zero at this point, which means that maximal destructive interference is not in fact obtained through LG violation here, but through the two contributing quasi-probabilities both being zero. However, this is an atypical point and the surrounding parameter ranges give LG violation and destructive interference. 

%[ELABORATE. MORE PLOTS?].

A question remains as to the absolute size of any of the LG violations obtained in this situation. The quasi-probability Eq.(\ref{quasi}) is bounded from below by $-\frac{1}{8}$, the L\"uders bound noted ealier \cite{HalTOA}.
To see how close the LG violations in the double-slit experiment are, we first normalize using a dichotomization and post-selection procedure.
We post-select onto a finite region of the screen, which is divided into two regions labeled by $s_2 = \pm 1$. Properly normalized quasi-probabilities $q(s_1, s_2)$ are then obtained from those obtained above by dividing by the probability of the finite region and all dependence on prefactor $N_\tau$ drops out. 
A convenient choice of post-selection region is to take $s_2 = + 1 $ to be $ -\pi / 2 < Y < \pi /2$ and $s_2 = -1 $ to be $ \pi/2 < Y < 3 \pi / 2$, i.e. the light and dark regions of the interference pattern where $\cos Y $ is positive or negative respectively.
We then readily find that the properly normalized post-selected quasi-probability is
\beq
q(s_1, s_2) = \frac{1}{4} \left( 1 + \cos (2 \phi) s_1 + \frac{2}{\pi} \sin (2 \phi)  s_2 \right).
\eeq
This has lower bound of about $ -0.05$  which is about 40\% of the L\"uders bound of $-0.125$.
(The normalized post-selected quasi-probability will still obey the L\"uders  bound, since it corresponds to restricting the initial state to lie in the subspace defined by the post-selection and the proof in Ref.\cite{HalTOA} still applies. The only difference is that it will generally not be possible to achieve the maximum.)

%\tb{
%Finally, note that it is natural to compare the LG approach, which assigns probabilities indirectly, with the different but related question as to the probability obtained using variable strength measurements which partially determine which slit the particle went through. This question has been considered by numerous authors  (see for example, Ref.\cite{WoZu}), and at least one experiment has been done \cite{Koc}. It can also be approached using the formalism developed here and earlier \cite{HalTOA} and yields modified versions of the quasi-probabilities obtained above.
%One can then analyze the deterioration of the interference pattern versus the certainty as to which slit the particle went through, and the analysis is very similar to that described in Ref.\cite{WoZu}.
%However, this approach is not a proper LG test since the measurements are now invasive, even if weak.
%}

\section{The Triple Slit Experiment}

We consider now the generalization of the approach described in Sections 2 and 3 to the triple-slit experiment \cite{Tri}.  We find what is new here is that there are three different types of interference terms.
There is again a discussion of the relationship between LG violation and destructive interference, but this relationship is more fluid than in the double-slit case, except for certain parameter values. More importantly, the triple-slit experiment provides testable examples of the new types of MR conditions that arise once we go beyond dichotomic variables to variables with three or more values \cite{HaMa2}.

\subsection{Interferences and quasi-probabilities}
% Note there is only one interference term in the DS case.

%\section{Triple-slit as a two time LG}

The formalism developed so far generalizes very readily to the triple slit. We choose the three slits to be located at $x = \pm L$ and $x=0$.
Alternatives at time $t_1$ are denoted by $n_1$ which may take values $-1, 0, 1$ and measurements at that time implemented through the projector $E_{n_1}$ and there is again a measurement $E_{n_2}$ at the screen. The two-time LG inequalities are simply
\beq
q(n_1, n_2) \ge 0
\eeq
where the quasi-probability $q(n_1, n_2)$ is the trivial generalization of Eq.(\ref{quasi}).
The two-time measurement probabilility $p_{12} (n_1, n_2)$ and quasi-probability $q(n_1, n_2)$ are again simply related and we have
\beq
q(n_1, n_2) = p_{12} (n_1, n_2)  +
 \sum_{\substack{n_1^\prime \\ n_1^\prime \ne n_1}} \Re D(n_1, n_2|n_1^\prime n_2).
\label{qpn}
\eeq
%where we have introduced the convenient condensed notation for the interference terms,
%\begin{equation}
%  I_{n_1, n_1^\prime} = 
%\label{ints}
%\end{equation}
The probability
at the second time can be written as
\begin{equation}
  p_2(n_2) = \sum_{n_1} p_{12}(n_1, n_2)  + \sum_{\substack{n_1, n_1^\prime \\ n_1 \neq n_1^\prime}}
\Re D(n_1, n_2|n_1^\prime n_2).
\end{equation}
In the double-slit case there was just one interference term. In the triple-slit case
there are three for fixed $n_2$.
% which we denote $I_{-,0}$, $I_{0,+}$, $I_{-,+}$.

We take the state just after the slits to be
\beq
| \psi \rangle = \sum_n \alpha_n | \psi_n \rangle,
\eeq
where $ | \psi_{\pm} \rangle $ are localized around $ x = \pm L$ and $| \psi_0 \rangle$ is localized around $x=0$. A convenient form for the coefficients $\alpha_n$, which ensures the proper normalization is
\begin{equation}
  \begin{split}
    \alpha_{+} &= e^{i\chi_+} \sin \theta \cos \phi, \\
    \alpha_{-} &= e^{i\chi_-} \sin \theta \sin \phi, \\
    \alpha_{0} &= \cos \theta.
  \end{split}
\end{equation}
Given this parametrisation the value of the screen probability density $p_2(x) = |\psi(x, \tau)|^2$ is found to be
\beq
  p_2(x) = |N_t|^2 \big[1 + 2  I_{0,+} + 2  I_{-,0} + 2  I_{-,+} \big],
\label{p2xtriple}
\eeq
where,
\begin{align}
 I_{0,+} &=  \sin \theta \cos \theta \cos \phi \cos(X_+),
\\
 I_{-,0} &=  \sin \theta \cos \theta \sin \phi \cos (X_-),
\\
 I_{-,+} &=\sin^2 \theta \cos \phi \sin \phi \cos(X),
\end{align}
and
\begin{align}
  X_{\pm} &= \frac{m}{2 \hbar \tau}(L^2 \mp 2Lx) + \chi_{\pm} \\
  X  &= \frac{2 m L x}{\hbar \tau} + \chi_- - \chi_+ = X_- - X_+.
\end{align}
The interference terms $I_{n_1, n_1'}$ are related to the off-diagonal terms of the decoherence functional in Eq.(\ref{qpn}) by
\beq
\Re D(n_1, n_2|n_1^\prime n_2) = |N_t|^2 I_{n_1, n_1'}.
\eeq
(And note that this notation differs from that in Ref.\cite{HaMa2} by a factor of $|N_t|^2$.)

The form of $p_{12}(n_1, x)$, using similar reasoning to the double-slit case, results in
\begin{equation}
 p_{12} (n_1, x) = |\alpha_{n_1}|^2 |N_t|^2.
\end{equation}
Calculating the quasi-probabilities for the triple slit with the $X_\pm$ substitution results in
\begin{align}
  q(+, x) &= |N_t|^2 \ \big[ \sin^2 \theta \cos^2 \phi + I_{0,+} + I_{-, +} \big], 
\label{q+x} \\
    q(-, x) &= |N_t|^2 \  \big[  \sin^2 \theta \sin^2 \phi + I_{-, 0} + I_{-,+} \big], 
\label{q-x}\\
  q(0, x) &= |N_t|^2 \ \big[ \cos^2 \theta +  I_{0,+} + I_{-,0} \big],
\label{q0x}
\end{align}
and note that the sum of these three yields $p_2(x)$ as expected. Parameters are readily found for which any of these three quasi-probabilities are negative. For example, consider $q(0,x)$ and take $X_+ = 0 = X_-$ and $\phi = 5 \pi / 4 $. Then $ q(0,x) = |N_t|^2 [ \cos^2 \theta - \sqrt{2} \sin \theta \cos  \theta ] $, which has minimum value $ |N_t|^2 ( 1 - \sqrt{3})/2 < 0$. Hence there are clear LG violations. (Like the double-slit case, we expect the violation to be a reasonable fraction of the maximum possible, but we do not compute this explicitly).

%\tb{LG violations?}

These three quasi-probabilities are determined experimentally using three dichtomizations of $n_1$ at $t_1$ and then using the formula Eq.(\ref{qMR}). For example, to determine $q(+,x)$, we consider the dichotomic variable $ \hat Q = 2 E_+ - \id $, and do non-invasive measurements to determine the sequential measurement probability 
$p_{12}^Q (s_1,x)$. This together with the probability $p_2 (x)$ permits the determinaton 
via Eq.(\ref{qMR}) of the quasi-probability $q(s_1, x)$, and setting $s_1 = +1$ yields the desired result. Similarly for the other two.

The screen probabilitity Eq.(\ref{p2xtriple}) involves a sum of all three interference terms but each of the three quasi-probabilities $q(n_1, x)$ involves only two interference terms. This means that, unlike the double-slit case, there is no general logical relationship between destructive interference and violation of the LG inequalities,
\beq
q(n_1, x ) \ge 0,
\eeq
and indeed they can behave quite independently. There are parameter ranges for which LG violation is accompanied by destructive interference, as in the double-slit case. There are paramater ranges which are the opposite to that case: LG violation but no destructive interference. The latter case arises because $p_2(x)$ is now a sum over three quasi-probabilities, two of which can cancel each other out if one is negative, leaving a third one which can exhibit either constructive or destructive inteference. It is also possible to find parameters for which LG violation and destuctive interference perfectly coincide.
%Some plots for a range of parameters are shown in Figure \ref{fig:triple}.
This lack of a general relationship is simply due to the fact that destructive interference alone for the triple slit experiment is a single and quite coarse-grained measure of the system's properties, compared to the more detailed description provided by the set of LG inequalities.

%[MORE EXAMPLES HERE? Figure removed since similar to later one.]

%\begin{figure}
%  \centering
% \includegraphics[width=0.6\linewidth]{fig_quasi_destructive_triple_slit}
 % \caption{The red region is the region of destructive interference and the grey region is the regions of negative quasiprobability for the case $X_+ = X_- = 0$. There is always a negative quasiprobability when destructive interference is present but not vice versa.}
%  \label{fig:triple}
%\end{figure}

\subsection{Two-time NSIT conditions in the Triple Slit Experiment}

More precise statements may be made by focusing in on NSIT conditions.
In Section 2 we noted that the NSIT condition Eq.(\ref{NSIT}) has a clearly logical connection to the two-time LG inequalities, namely the former implies the latter, but not conversely. This is because the NSIT condition implies that the single interference term encountered there must vanish but the LG inequalities require only that it is not too large.

However, this was for measurement of a single dichotomic variable at the first time. For the triple-slit experiment, in which we are in effect measuring a three-valued variable at the first time,  a more complicated relationship arises. For a system taking three values at the first time, the natural analogue of the NSIT condition Eq.(\ref{NSIT}) is,
\begin{equation}
  \sum_{n_1} p_{12}(n_1, n_2) = p_2(n_2).
  \label{NSIT3}
\end{equation}
We then readily see from the above that in the triple-slit experiment this means that the sum of all three interference terms must vanish,
\begin{equation}
  I_{-, 0} + I_{0, +} + I_{-, +} = 0. \label{I123}
\end{equation}
That is, there is neither constructive nor destructive interference. However, this clearly does not imply that all the LG inequalities are satisfied, since this requires that the sums of pairs of interferences terms are sufficiently small.

\begin{figure}
  \centering
  \includegraphics[width=.7\linewidth]{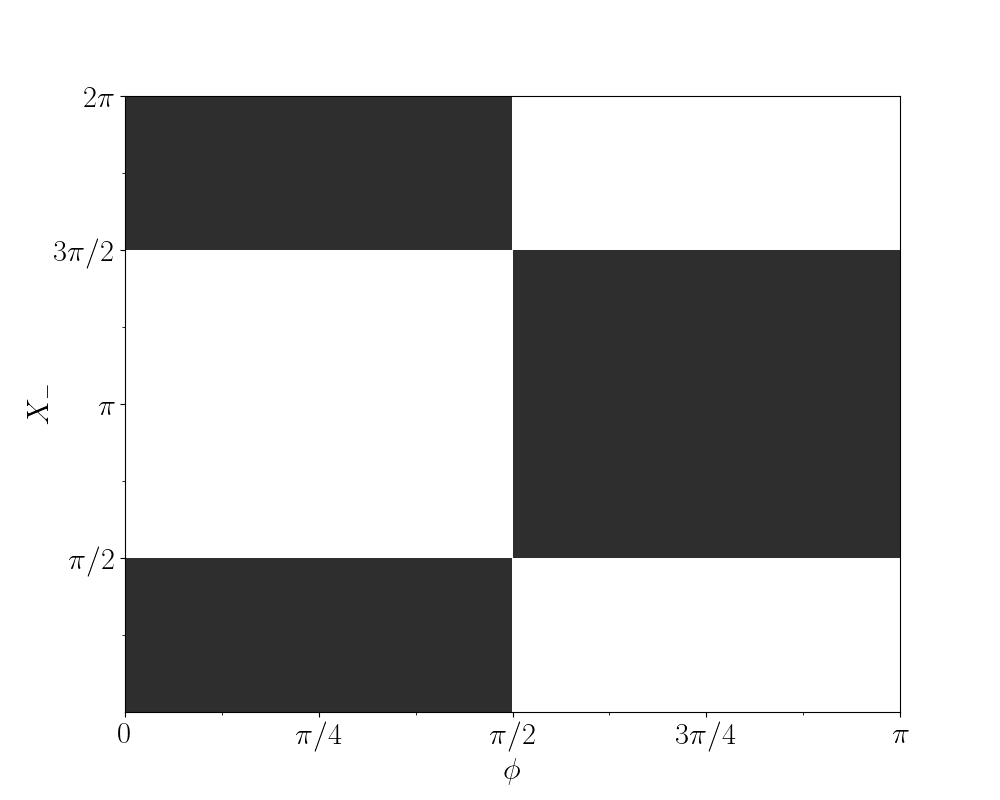}
  \caption{The shaded region of this parameter space shows the regions where the quasiprobabilities
  are negative for the triple slit whilst the overall NSIT condition is met. In this figure $X_+ =
  0.001$. Hence LG violations are still possible when the overall NSIT condition is met. }
  \label{fig:nsit}
\end{figure}

It is of interest to illustrate the parameter ranges that there is LG violation but with the NSIT conditions Eq.(\ref{NSIT3}) satisfied. The NSIT condition may be solved
%written explicitly as,
%\begin{equation}
%  \cos \theta \sin \theta \cos \phi \cos(X_+) + \cos \theta \sin \theta \sin \phi \cos(X_-)  +
%  \sin^2 \theta \sin \phi \cos \phi \cos(X) = 0.
%\label{eq:nsit_calculated}
%\end{equation}
%This condition can be rearranged as follows
to yield
\begin{equation}
  \tan \theta = - \frac{\cos \phi \cos (X_+) + \sin \phi \cos (X_-)}{\sin \phi \cos \phi \cos (X)}
  \label{eq:tan_theta},
\end{equation}
which can be readily substituted into the forms for the quasiprobabilities, Eqs.(\ref{q+x})--(\ref{q0x}).
%This results in quasiprobabilities of the form
%\begin{align}
 % q(+, x) &\propto \sin^2 \theta \Bigg(\cos^2 \phi + \cos \phi \sin \phi \cos(X) \Bigg(1 -
 %   \frac{1}{1+\tan \phi \ \frac{\cos (X_-)}{\cos (X_+)}}\Bigg)\Bigg) \\
 % q(-, x) &\propto \sin^2 \theta \Bigg(\sin^2 \phi + \cos \phi \sin \phi \cos(X) \Bigg(1 -
 %   \frac{1}{1+\cot \phi \ \frac{\cos (X_+)}{\cos (X_-)}}\Bigg)\Bigg) \\
 % q(0, x) &\propto \cos^2 \theta \Big(1 - \frac{[\cos \phi \cos( X_+)+ \sin \phi \cos
 % (X_-)]^2}{\sin \phi \cos \phi \cos(X)}\Big).
 % \label{eq:nsit_lg_violation}
%\end{align}
The parameter ranges for which at least one of these three quasi-probabilities is negative are shown in Figure \ref{fig:nsit}.
There are clearly substantial regions of LG violation even though the NSIT condition is satisfied.

Examples of this form, in which LG violation is observed when a NSIT condition holds, have been discovered previously and observed experimentally \cite{Ema,Ema2,Geo}.
A general analysis of this initially surprising phenomemon was given in Ref.\cite{HaMa2}.  The key point is that for situations such as the triple-slit experiment, there is not just one NSIT condition. The other NSIT conditions  may be found by considering all possible dichotomic variables at the first time, which we denote $\hat Q(n_1)$ and are given by
\beq
\hat Q(n_1) = 2 E_{n_1} - \id,
\eeq
where $\id$ denotes the identity operator.
Each choice of $\hat Q$ produces a two-time probability $p_{12}^Q(s_1, n_2)$ for $s_1 = \pm 1$ and if we require that each satisfies its own NSIT condition,
\begin{equation}
  \sum_{s_1} p_{12}^Q(s_1, n_2) = p_2(n_2),
  \label{eq:nsit_many_dich}
\end{equation}
we then find that the interference terms must satisfy,
\begin{align}
  I_{0, +} + I_{-, +} &= 0, \label{I1} \\
  I_{-, 0} + I_{-, +} &= 0, \label {I2} \\
  I_{-, 0} + I_{0, +} &= 0, \label{I3}
\end{align}
which actually imply Eq.(\ref{I123}). So there are in fact three independent NSIT conditions (for fixed $n_2$) and if any three of the conditions Eq.(\ref{I123}), Eq.(\ref{I1})-(\ref{I3}) are satisfied than all interferences terms are zero and all NSIT conditions hold. This means that if {\it all} NSIT conditions hold then the LG inequalities must be satisfied, in parallel with the double-slit case, but if only some NSIT conditions hold, then some of the LG inequalities can still be violated.

\subsection{Consequences of von Neumann measurements}

Another new feature of the systems with three or more levels, compared to the dichotomic case, is the possibility of measuring the correlator in more than one way, all macrorealistically equivalent, but different in quantum theory. The discussion so far involved determining the correlator Eq.(\ref{corr}) through a measurement of the dichotomic variable $\hat Q$, a L\"uders measurement. However, for a system with three or more levels we can instead do a von Neumann measurement, using a projector $E_n$ to determine which level the system is, and then construct the correlator from the joint probability $p_{12} (n_1, n_2)$. The resulting correlator $C_{12}^{vN}$ differs from the L\"uders one, by an interference term describing the interference between the states in the degenerate subspace of $\hat Q$. One consequence of this is that is that LG inequalities constructed from the von Neumann correlators can violate the L\"uders bound \cite{Dak,deg,EmaExp,PQK,KQP}, which recall, is $-\frac{1}{8}$ for the two-time quasi-probabilities. (See Ref.\cite{HaMa2} for more details on the above). 
A L\"uders violation may be demonstrated in the triple-slit experiment but this is a lengthy calculaiton and will not be given here.

Instead, we briefly describe a different consequence of using the von Neumann correlators. We showed in Section 4(B) that it is possible to satisfy the overall NSIT condition Eq.(\ref{NSIT3}) but violate at least one LG inequality. This cannot happen when the NSIT condition is the condition Eq.(\ref{eq:nsit_many_dich}) specific to the variable $\hat Q$ appearing in the LG inequality.
However, this conclusion changes when the correlator in the LG inequality is replaced with the von Neumann one. 
%\beq
%1 + \langle Q_1 \rangle + \langle Q_2 \rangle + C_{12}^{vN} \ge 0,
%\eeq
This replacement may
be shown to be equivalent to replacing, for example, the quasi-probability Eq.(\ref{q+x}), with 
\beq
q^{vN} (+, x) = |N_t|^2 \ \big[ \sin^2 \theta \cos^2 \phi + I_{0,+} + I_{-, +} - I_{-,0}  \big].
\label{qvn}
\eeq
(See for example Eq.(6.10) in Ref.\cite{HaMa2}. Note also that a construction of the general form
Eq.(\ref{qMR} can yield the same answer).
The NSIT condition associated with $\hat Q$ is satisfied when $ I_{0,+} + I_{-,+} = 0$,
which is easily achieved by taking $ \phi = \pi/2$.  We then have
\beq
q^{vN} (+, x) = |N_t|^2 \ \big[ \sin^2 \theta \cos^2 \phi  -  \sin \theta \cos \theta \sin \phi \cos (X_-) \big], 
\label{qvn2}
\eeq
where we have inserted the explicit value of $I_{-,0}$. Violations of $q^{vN}(+,x) \ge 0 $ are then easily found. For example, $\phi = \pi/2$, $ \theta = \pi/4$ and $X_- = 0$ yields
$
q^{vN} (+,x) = - |N_t|^2/2 < 0.
$
Hence we have an example of an LG inequality for a single variable $\hat Q$ that can be violated even when the NSIT condition for $\hat Q$ is satisfied, a striking contrast with the dichotomic case where this is impossible.
A very similar feature in a multi-level system was noted in Ref.\cite{KQP}.

\section{Leggett-Garg Violations in the simple harmonic oscillator with a single coherent state}

Now we shift our analysis from interference experiments to investigating another common type of
experimental setup used in quantum coherence experiments, namely the coherent state of a harmonic
oscillator. This work is inspired by a recent discussion by Bose et al
\cite{Bose} which suggests that the ``classical-like'' coherent state of the harmonic oscillator
may still exhibit a significant violation of the LG inequalities. (See also Ref.\cite{Myung}).
This type of model describes a
number of macroscopic oscillator systems that can be realized experimentally, and hence provides a 
possible path to LG tests on truly macroscopic systems. Bose et al considered the four-time LG inequalities for the coherent state and used numerical methods to calculate the correlators and exhibit a LG violation. Here we take the simpler case of a two-time LG inequality which confers the advantage that it will be considerably easier to measure experimentally, involving just one correlator, rather than four. Furthemore, a violation of a two-time LG inequality renders redundant further checks involving LG conditions at more time since macrorealism has already failed \cite{HalLG4}.
Although the correlators for this system are considerably harder to evaluate than those for simple spin systems, we are able to find exact analytic expressions. We will exhibit a violation of the two-time inequality.

We suppose that measurements are made which determine whether the particle is in $x<0$ and $x>0$ at each time and the dichotomic variable is taken to be $\hat Q = P_+ - P_-$ where the projectors are $P_{\pm} = \theta(  \pm \hat x) $. We focus on the quasi-probability $q(-,+)$ which may be written,
\begin{align}
  q(-, + ) &= {\rm Re} \langle \psi | e^{iHt} P_+ e^{-iHt} P_- | \psi \rangle \\ &= {\rm Re} \int_{0}^\infty dx \
  \psi^* (x, t) \langle x | e^{-iHt} P_- |\psi \rangle,
  \label{eq:quasi_sho}
\end{align}
where the initial state is taken to be the gaussian,
\beq
\psi(y) = N_s \exp \left( - \frac{y^2}{4\sigma^2} + \frac{ip_0 y }{\hbar} \right), \label{eq:sho_initial}
\eeq
where
\beq
N_s =
\frac{1}{(2\pi \sigma^2 ) ^{1/4}}.
\eeq
The harmonic oscillator propagator is,
\begin{equation}
\langle x | e^{-iHt} P_- | y \rangle = N_p  \exp \left( \frac{  i m \omega}{2\hbar \sin \omega t } \left[ \left(x^2 + y^2 \right) \cos \omega t - 2xy \right] \right), 
\eeq
where
\beq
 N_p = \left( \frac{m\omega}{2\pi i \hbar \sin \omega t } \right)^{1/2}.
\eeq
We therefore get
\begin{align}
  \langle x | e^{-iHt} P_- | \psi \rangle &= \int_{-\infty }^0 dy \langle x | e^{-iHt} | y \rangle
  \psi(y) \\ &= {\rm Re} \left\lbrace N_s N_p \int^0_{-\infty}dy \exp\left(-ay^2 + iby\right) \exp \left(
  \frac{im\omega x^2}{2\hbar \sin \omega t} \cos \omega t \right) \right\rbrace
\end{align}
where
\begin{equation}
    a =  \left( \frac{m\omega \cos \omega t}{2i \hbar \sin \omega t} + \frac{1}{4\sigma^2}
    \right),  \qquad b=  \left(  \frac{p_0}{\hbar} -\frac{m \omega x}{\hbar \sin \omega t} \right).
\end{equation}
Evaluating the integral we obtain
\begin{align}
 \langle x | e^{-iHt} P_- | \psi \rangle = {\rm Re} \left\lbrace N_s N_p \frac{1}{2} \sqrt{\frac{\pi}{a}}
 e^{-b^2/4a} \left(1-i \text{erfi}\left(\frac{b}{2\sqrt{a}}\right)\right) \exp \left(
 \frac{im\omega x^2}{2\hbar \sin \omega t} \cos \omega t \right) \right\rbrace,
\end{align}
where $\text{erfi}$ is the imaginary error function.
The quasi-probability is therefore given by
\begin{equation}
q(-,+) = {\rm Re} \left\lbrace |N_s|^2 |N_p|^2 \frac{\pi}{2|a|}  \int^\infty_0 dx \ e^{-A z^2(x)} \text{erfc} \left(i z(x) \right)  \right\rbrace,
\end{equation}
where ${\rm erfc}$ is the complementary error function and we have
\begin{align}
z(x) &= \frac{b}{2\sqrt{a}} = \frac{1}{2\sqrt{a}}
\left(  \frac{p_0}{\hbar} -\frac{m \omega x}{\hbar \sin \omega t} \right)  ,
\\
A &= a \left( \frac{1}{a} + \frac{1}{a^*} \right).
\end{align}
Performing a simple change of variables we arrive at a form,
\begin{equation}
q(-,+) = {\rm Re} \left\lbrace |N_s|^2 |N_p|^2 \frac{\pi \hbar  \sqrt{a}  \sin \omega t}{|a| m \omega} \int_{-\infty}^{z(0)} dz e^{-A z^2} \text{erfc} \left(i z \right)  \right\rbrace.
\end{equation}
The remaining integral can be evaluated with the result,
\begin{equation}
q(-, +) = \frac{1}{4} \left[ 1 +  \text{erf} \left(\sqrt{A} z(0) \right) +  4\ {\rm Re} \left\lbrace T
\left[ \sqrt{2A} z(0), \frac{i}{\sqrt{A}} \right]  \right\rbrace  \right].
\label{eq:quasi_sho_full}
\end{equation}
Here, $ T[h,a]$ denotes the Owen T-function \cite{Owen},
\beq
T[h,a] = \frac{1}{2 \pi} \int_0^a \ dx \ \frac{ \exp( - \frac{1}{2} h^2 (1 + x^2)) }  { 1 + x^2 }.
\eeq
The other three components of the quasi-probability are obtained using simple symmetry arguments applied to the above calculation and we find the result,
\beq
q(s_1, s_2) = \frac{1}{4} \left[ 1 + s_2  \text{erf} \left(\sqrt{A} z(0) \right) -   4 s_1 s_2 {\rm Re} \left\lbrace T
\left[ \sqrt{2A} z(0), \frac{i}{\sqrt{A}} \right]  \right\rbrace  \right].
\eeq
From this we may read off the averages $\langle \hat Q_1 \rangle = 0 $,
\beq
\langle \hat Q_2 \rangle  =  \text{erf} \left(\sqrt{A} z(0) \right),
\eeq
and the correlator,
\beq
C_{12} = -4 \ {\rm Re} \left\lbrace T
\left[ \sqrt{2A} z(0), \frac{i}{\sqrt{A}} \right]  \right\rbrace.
\label{c12SHO}
\eeq

The arguments of the error function and T-function have the explicit expressions, 
\begin{align}
 \sqrt{A} z(0) &= \frac{  \sqrt{2} p_0 \sigma}{ \hbar } \left( 1 + 4 \omega^2 t_s^2  \cot^2 (\omega t)\right)^{-\frac{1}{2}},
\label{6a}
\\
\frac{i}{ \sqrt{A} } &= \frac{i} {\sqrt{2} } \left( 1 + 2 i \omega t_s \cot (\omega t) \right)^{1/2}.
\label{6b}
\end{align}
where $ t_s = m \sigma^2 / \hbar $ (and is the wave packet spreading timescale for the free particle).
For the simple harmonic oscillator coherent state we have $\sigma^2 = \hbar / (2 m \omega) $ and it follows that $2 \omega t_s  = 1$ and the above expressions simplify to
\begin{align}
 \sqrt{A} z(0) &= \frac{  \sqrt{2} p_0 \sigma}{ \hbar } \left( 1 +  \cot^2 (\omega t)\right)^{-\frac{1}{2}},
\\
\frac{i}{ \sqrt{A} } &= \frac{i} {\sqrt{2} } \left( 1 +  i\cot (\omega t) \right)^{1/2}.
\end{align}
The free particle case is obtained by taking 
the limit $\omega \rightarrow 0 $ in Eqs.(\ref{6a}), (\ref{6b}),
and we have
\begin{align}
 \sqrt{A} z(0) &= \frac{  \sqrt{2} p_0 \sigma}{ \hbar } \left( 1 + \frac{1} { \tau^2} \right)^{-\frac{1}{2}},
\\
\frac{i}{ \sqrt{A} } &= \frac{i} {\sqrt{2} } \left( 1 + \frac{i}{ \tau} \right)^{1/2},
\end{align}
where $\tau =  t / (2 t_s)$. This therefore has exactly the same form as the simple harmonic oscillator case with $ \tan(\omega t)$ replaced by $\tau$, and therefore the plots of each are identical with the appropriate parameterization. 

A plot of the quasi-probability  $q(-,+)$ is shown in Figure \ref{fig:free_particle_negative_p}. It shows negativity (LG violation) for a range of $\tau$, although the negativity is quite small, about 10\% of the maximum negativity in the quasi-probability of $-0.125$. The other three quasi-probabilities exhibit no negativity and also show the expected quasi-classical behaviour.

\begin{figure}
  \centering
  \includegraphics[width=.7\linewidth]{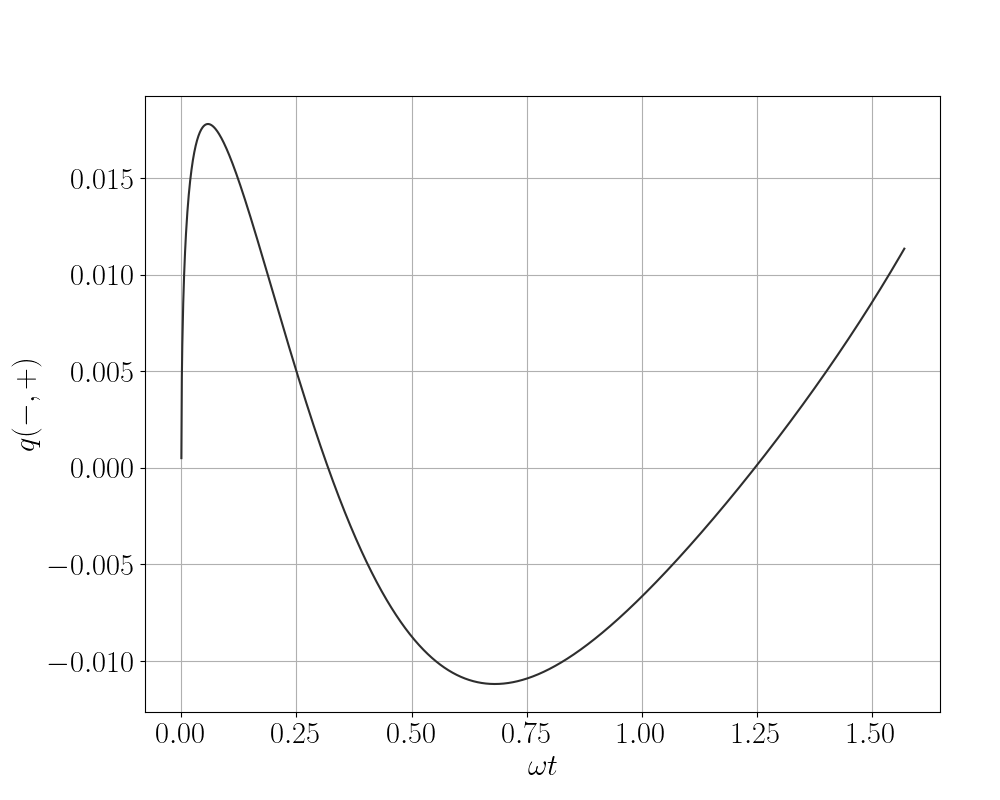}
  \caption{The quasi-probability $q(-, +)$ for the simple harmonic oscillator as a function of $\omega t$ for the case $p' = p_0 \sigma / \hbar =-1$. It shows a region of negativity with lowest value approximately $-0.011$.
}
  \label{fig:free_particle_negative_p}
\end{figure}

%It is also of interest to see the form of the correlator Eq.(\ref{c12SHO}) and this is shown in Figure \ref{fig:corr} for fixed momentum. It becomes close to a pair of straight lines for large $p$, which is what one would expect on classical grounds. 

Another natural situation to examine using the correlator Eq.(\ref{c12SHO}) would be a set of three-time or four-time LG inequalities. However, the above calculation leading to the correlator Eq.(\ref{c12SHO}) is only relevant to situations in which the system is in a state of the form Eq.(\ref{eq:sho_initial}) at the beginning or end of each time interval which we have found is too restrictive to yield any non-trivial results.

%\begin{figure}
 % \centering
  %\includegraphics[width=.7\linewidth]{fig9}
 % \caption{The correlator $C_{12}$ for the simple harmonic oscillator as a function of $\omega t$, with $p' = -1$.
%}
 % \label{fig:corr}
%\end{figure}

We make some final comments to contrast the LG violations for the simple harmonic oscillator in a gaussian state with the LG violations in interference experiments. The quasi-probability $q(s_1, s_2)$ of the form Eq.(\ref{quasi}) may be conveniently written using the Wigner representation \cite{Wig} as
\beq
q(s_1, s_2) = 2 \pi \hbar \int  dX dp \  W_{s_1 s_2} (X,p) W_\rho (X,p),
\eeq
as discussed in Ref.\cite{HalTOA}.
Here, $W_\rho (X,p)$ is the Wigner transform of the initial state $\rho$, 
%\beq
%W_\rho (X,p) = \frac{1}{2 \pi \hbar} \int_{-\infty}^{\infty} d \xi \exp \left( - \frac{i}{\hbar} p \xi \right) \ \rho ( X + \frac{1}{2}\xi, X - \frac{1}{2} \xi ),
%\eeq
and $W_{s_1 s_2} (X,p)$ is the same transform of the operator part of Eq.(\ref{quasi}).
%\beq
%A_{s_1 s_2} = \frac{1}{2} \left( P_{s_1} (t_1) P_{s_2} (t_2) + P_{s_2} (t_2) P_{s_1} (t_1)\right).
%\eeq
%The quantities $W_{s_1 s_2}(X,p)$ and $W_{\rho} (X,p)$ tend to be largely positive in phase space, with occasional oscillations to negative values.  There are therefore two distinct ways in which $q(s_1,s_2)$ may become negative and we get a two-time LG violation -- it can come predominantly from the state or predominantly from the operator $A_{s_1 s_2}$. 
For the superposition states considered in the interference experiments the Wigner functions are negative and this is the primary source of the LG violation (noting also that the essentially semi-classical approximations used will probably render $W_{s_1 s_2} (X,p)$ non-negative).
For the gaussian state by contrast, the Wigner function is always non-negative, hence for the harmonic oscillator case considered in this section, the LG violation comes from the negativity of $W_{s_1 s_2} (X,p)$ (whose explicit form is given in Ref.\cite{HalTOA} for the free particle case).
Hence the LG violations in each case have quite different origins. 

%Note also that in the latter case, the negativity of $W_{s_1 s_2} (X,p)$ is related to the sharpness of the projectors. In a realistic experimental situation the measurements will correspond to smoothed out projectors and one would expect the negativity of $W_{s_1 s_2} (X,p)$ and the subsequent LG violation to be lessened, so a more detailed analysis is required to estimate how smooth the projectors can be whilst still maintaining a LG violation. This will be explored in more detail elsewhere.

\section{Summary}

We have explored LG tests for macrorealism in a number of systems described by continuous variables: the double-slit experiment, the triple-slit experiment and the free particle and simple harmonic oscillator in a gaussian state. These systems are of particular interest in the drive to develop LG tests for progressively larger systems with some claim towards being macroscopic.
The main result of this paper is that regimes in which LG inequalities are robustly violated are readily found in all cases.

The second main aim of the paper concerned the relationship between destrutive interference and LG violation.
In the double-slit case described in Section 3, we found that LG violations are essentially always accompanied by destructive interference, illustrating a more general result shown in Section 2.
The converse is not true in general, except for special parameter values. However, the fact that destructive interference does not always imply LG violation means that there are situations in which one {\it can} in fact assign probabilities to the paths in an interferometer, using an indirect procedure, even when destructive interference is present -- i.e. there is in some circumstances an underlying classical model of a situation commonly thought of as ``quantum''.
Differently put, destructive interference alone is not a decisive measure of quantumness.
We compared our results with the earlier work involving the MZ interferometer \cite{Pan,Robens,KoBr}, where the relationship is different.

The triple-slit experiment revealed a parameter-dependent relationship between destructive interference and LG violation, due to the fact that it involves three independent interference terms, rather than just the one present in the double-slit case. In fact parameters were readily found such that LG violations could arise without any destructive interference being present. 
The triple-slit experiment also permits new possibilities not present for measurements of dichotomic variables, namely situations in which there are LG violations but with certain NSIT conditions satisfied. We confirmed that parameters exist for which this possibility arises.
We also briefly discusssed L\"uders bound violations in the triple-slit-experiment.
These features illustrate in a specific model the general novel properties of many-valued systems described in Ref.\cite{HaMa2}.

In these interference experiments, LG violations arose due to the presence of superposition states in position, i.e. from negative Wigner function. By contrast, the gaussian initial state used in the simple harmonic oscillator case has non-negative Wigner function, but violations of a two-time LG inequality are still possible. This provides an analytically tractable example of a LG violation
simpler and easier to measure than that given in Ref.\cite{Bose}.
%Furthermore, the simplicity of the two-time inequality compared to the four-time one considered in Ref.\cite{Bose} is an advantage in terms of ease of measurement experimentally.
A future paper will explore LG tests for a variety of experimentally accessible states of the simple harmonic oscillator in considerably more detail \cite{HaMa3}.

\section{Acknowledgements}

We are grateful to Sougato Bose, Clive Emary, Dipankar
Home, George Knee, Johannes Kofler, Raymond Laflamme, Seth Lloyd,
Shayan Majidy, Owen Maroney, Clement Mawby, Alok Pan
and James Yearsley for many useful discussions and email
exchanges about the Leggett-Garg inequalities over a long
period of time.

%We are grateful to Alok Pan for some useful communications.

%Gonzalo Muga and James Yearsley for many useful conversations on these topics over a long period of time.

%\input refs.tex

\end{document}